# On evolutionary games with periodic payoffs

E. Ahmed and Muntaser Safan
Mathematics department, Faculty of Science, Mansoura 35516, EGYPT.

Abstract: Two cases of evolutionary stable strategy with periodic payoffs are studied. The first is a generalization of Uyttendaele et al. The second is prisoner's dilemma with periodic payoff. It is shown that reducing the defection payoff by a periodic term is sufficient to introduce cooperation provided that the initial fraction of cooperators is greater than 0.6.

1. Introduction:
Seasonality is an important phenomena in biology. Yet studying evolutionary games [Hofbauer and Sigmund 1998 and Maynard Smith and Price 1978] with periodic payoffs is still in its infancy [Uyttendaele et al 2012 and Broom 2005].

In sec.2 the work of Uyttendaele et al will be generalized to the case $o \ll 1$ and $\alpha = 1$ which has not been studied. In sec.3 the case of prisoner's dilemma game with periodic payoff will be introduced.

Sec.2 generalized Uyttendael et al case:
Uyttendaele et al have studied evolutionary game with payoff matrix

$$A = \begin{bmatrix} 0 & 0 & 2+o\cos(\rho t) \\ \alpha & 0 & \alpha \\ 2-o\cos(\rho t) & 0 & 0 \end{bmatrix} \quad (1)$$

numerically. The replictor dynamics equations are

$$dp_i/dt = p_i[(A\begin{bmatrix} p_1 \\ p_2 \\ p_3 \end{bmatrix})_i - p^t A p], i=1,2,3 \quad (2)$$

$p_1 + p_2 + p_3 = 1$

We will study the case $o \ll 1$ which was not considered in their paper. The equilibrium solution implies:

$\alpha = 2p_3/(p_1+p_3), 2p_1+p_2=1 \Rightarrow p_1=p, p_3=p, p_2=1-2p, \alpha=1 \quad (3)$

To study the time dependent case we set

$p_1 = p_3 = p_0 + ox(t), p_2 = 1 - 2p_0 - 2ox(t) \quad (4)$

and linearize in $o$. We get

$$dx/dt = p_0^2 \cos(\rho t),$$
$$0 < p_0 < \min\{1/2, \rho/o\}. \quad (5)$$

Hence

$$p_1 \approx p_3 \approx p_0 + (op_0^2/\rho)\sin(\rho t) + O(o^2) \quad (6)$$
$$p_2 \approx 1 - 2p_0 - 2(op_0^2/\rho)\sin(\rho t) + O(o^2)$$

In this case all strategies co-exist.

Sec.3 Prisoner's Dilemma game with periodic payoff:

Prisoner's dilemma game is a classic model for cooperation between selfish individuals [Hofbauer and Sigmund 1998]. Its payoff matrix is given by

$$A = \begin{bmatrix} R & S \\ T & U \end{bmatrix}, T > R > U > S, 2R > T + S \quad (7)$$

The evolutionary stable strategy is defect (p=0). The reason for this negative result is the condition U>S. This negative result has been corrected by the following mechanisms: Relatedness, memory and inhomogeneity [Shehata 2011].

Here we ask the following question: If a periodic term is added to reduce U will this introduce a kind of cooperation? Consider PD game with the following payoff matrix

$$A = \begin{bmatrix} R & 0 \\ T & U_0 - U_1\cos(t) \end{bmatrix}, U_1 > U_0 \quad (8)$$

The replicator dynamics equation is

$$dp/dt = p(1-p)[p(R - T + U_0 - U_1\cos(t)) - U_0 + U_1\cos(t)] \quad (9)$$

Solving (9) numerically we found that the final result crucially depend on the initial fraction of cooperators. If it was greater than 0.6 then cooperation dominates otherwise defect is the evolutionary stable strategy as shown in the figure.

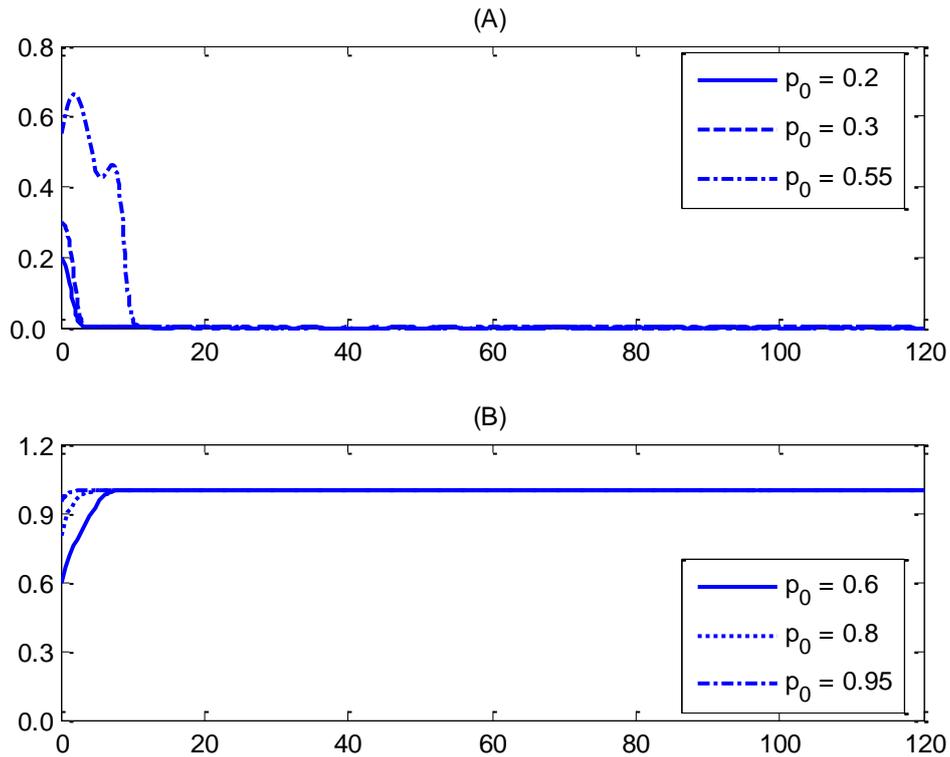